\documentclass[aps,prb,showpacs,showkeys,groupedaddress,twocolumn]{revtex4}%
\usepackage{amsfonts}
\usepackage{amsmath}
\usepackage{amssymb}
\usepackage{graphicx}
\usepackage{graphics}

\begin{document}

\title{Spin Nernst effect in the absence of a magnetic field}
\author{Xuele Liu$^1$ and X.C. Xie$^{1,2}$}

\address{
$^1$Department of Physics, Oklahoma State University, Stillwater,
Oklahoma 74078\\
$^2$Institute of Physics, Chinese Academy of Sciences, Beijing
100190, China}

\begin{abstract}
We study the spin Nernst effect of a mesoscopic four-terminal
cross-bar device with the Rashba spin-orbit interaction (SOI) in the
absence of a magnetic field. The interplay between the spin Nernst
effect and the seebeck coefficient is investigated for a wide range
of the Rashba SOI. When no peaks appeared in the seebeck
coefficient, an oscillatory spin Nernst effect still occurs. In
addition, the disorder effect on the spin Nernst effect is also
studied. We find that the spin Nernst effect can be enhanced up to
three-fold by disorder. Besides, due to the interface effect, the
counter-propagating of the charge current to the direction of the
temperature gradient is possible for a nonuniform system.
\end{abstract}
\keywords{A. Rashba spin-orbit interaction; C. Disorder; D. Spin
Nernst Effect; D. seebeck coefficient}

\pacs{72.15.Jf,72.25.-b, 73.23.-b, 73.43.-f}

\maketitle

With the development of the micro-fabrication technology and the
low-temperature measurement technology, a great amount of efforts
have been paid for the
research of the thermoelectric properties in the last two decades \cite%
{ref10,ref11}. Comparing to the conductance, the thermoelectric coefficients
of electronic systems are more sensitive to the details of the density of
states\cite{ref7,ref8,ref9}, which is very important for the design of the
electronic devices. The thermopower (seebeck coefficient) of the quantum dot
was measured in the last few years\cite{ref10}. Recently, the Nernst effect,
a Hall-like thermal effect, has been theoretically studied \cite{ref14} and
had been detected, for example, in bismuth\cite{ref13} in which, with the
existence of a perpendicular magnetic field, a transverse current is induced
by the longitudinal thermal gradient.

In the spintronics area, the spin thermal coefficients are also of
focus recently\cite{ref7,ref8,ref9}. In a recent paper, by
considering a system with a spin-orbit interaction (SOI), the Nernst
effect and a novel thermal effect, the spin Nernst effect, have been
fully studied in a two-dimensional electron gas\cite{ref0}. It is
found that, because of a perpendicular magnetic field $B$, the
Nernst signal exhibits a series of peaks. When the SOI exists, the
peaks split and the spin Nernst effect appears. With a small $B$ or
a large SOI, the spin Nernst effect becomes more pronounced. It also
shows that the spin Nernst effect is easier to be affected by
disorder than the Nernst effect.

There is no doubt that a perpendicular magnetic field $B$ is essential for
the existence of the Nernst effect. However, in the spin Hall effect, the
transverse spin current is due to a SOI rather than a perpendicular magnetic
field. Similarly, for the spin Nernst effect, $B$ may not be needed either.
One may suspect that the spin Nernst effect is in fact the combination of
the existence of thermopower and a SOI. Thus, the focus of the current work
is to study the spin Nernst effect in the absence of a perpendicular
magnetic field, and its interplay with the thermopower.

In this paper, the property of spin Nernst effect is developed in a
two-dimensional electron gas system with a Rashba SOI but without a
perpendicular magnetic field $B$. For this set-up, the Nernst effect
disappears thus we focus on the spin Nernst effect -- a transverse spin
current induced by a longitudinal thermal gradient $\Delta \mathcal{T}$. A
traditional way to analyze such a Hall-like system is to add vertical probes
to detect the transverse properties. Thus we set a four-terminal cross-bar
sample, as shown in Fig.1\cite{ref0}. A longitudinal thermal gradient $%
\Delta \mathcal{T}$ is added between the leads 1 and 3. This thermal
gradient induces a transverse spin current $J_{s}$ in the closed boundary
condition with a SOI, which can be measured at leads 2 and 4. The seebeck
coefficient of such a system can be directly measured at leads 1 and 3.

By using a tight-binding model and the Landauer-Buttiker (LB)
formula, the spin Nernst coefficient $N_{s}$ ($N_{s}\equiv
J_{s}/\Delta \mathcal{T}$) and the seebeck coefficient $S$ ($S\equiv
-\Delta V/\Delta \mathcal{T}$) are calculated. The Rashba SOI used
in our calculations covers a wide range with some beyond the
accessibility of today's sample. The seebeck coefficient $S$ shows a
few peaks consequently when the fermi energy $E_{F}$ goes through
the energy band. Due to the interface of our setting (zero Rashba
SOI at lead 2,4), we find a negative $S$. It is confirmed that spin
Nernst effect can not be simply thought as the combination of the
seebeck coefficient and the Spin hall effect\cite{ref0}. A big spin
Nernst coefficient $N_{s}$ can be found with a zero seebeck
coefficient $S$. However, when the peaks of seebeck coefficient
occur with a non-zero Rashba SOI, the spin Nernst effect exhibits
big amplitude or sometimes also peaks. The Fermi energy $E_{F}$ also
affects $N_{s}$. When the Fermi energy $E_{F}$ is close to the
bottom of the energy band ($-4t$), the oscillatory amplitude of
$N_{s}$ becomes more pronounced. The effect of disorder on $N_{s}$
is also investigated. When $E_{F}=-3.8t$, we can see a large
increase of $N_{s}$ with increasing of the strength of disorder. Its
value at the peak is about three-fold of that
without disorder. In addition, we find that the strength of disorder when $%
N_{s}$ vanishes, indicating that the system goes into an insulating regime,
is independent of the Fermi energy.

\begin{figure}[tbp]
\includegraphics[bb=146 485 391 674,scale=.7,clip=]{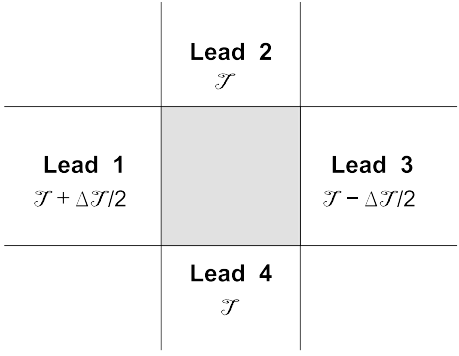}
\caption{Schematic diagram of the four-terminal cross-bar sample. The area
with SOI is marked by gray. A thermal gradient $\Delta \mathcal{T}$ is
applied between the longitudinal lead-$1$ and lead-$3$.}
\end{figure}

In the tight-binding representation, the Hamiltonian with SOI can be written
as:\cite{ref15},
\begin{eqnarray}
H &=&\sum_{\mathbf{i}\sigma }\varepsilon _{\mathbf{i}}c_{\mathbf{i}\sigma
}^{\dag }c_{\mathbf{i}\sigma }+\sum_{\mathbf{i}\sigma \sigma ^{\prime }}[c_{%
\mathbf{i}+\delta y,\sigma }^{\dag }(-t\mathbf{I}-i\sigma _{x}V_{R})_{\sigma
\sigma ^{\prime }}c_{\mathbf{i}\sigma ^{\prime }}  \notag \\
&&+c_{\mathbf{i}+\delta x,\sigma }^{\dag }(-t\mathbf{I}+i\sigma
_{y}V_{R})_{\sigma \sigma ^{\prime }}c_{\mathbf{i}\sigma ^{\prime }}+H.c.]
\end{eqnarray}%
where $c_{\mathbf{i}\sigma }^{\dag }$($c_{\mathbf{i}\sigma }$) is the
creation (annihilation) operator of electrons in the site $\mathbf{i}=(n,m)$
with spin $\sigma $, and $\delta x$ and $\delta y$ are the unit vectors
along the x and y directions. $\varepsilon _{\mathbf{i}}$ is the on-site
energy, which is set to $0$ everywhere for the clean system. When the center
region is a disorder system, $\varepsilon _{\mathbf{i}}$ is set by a uniform
random distribution [-W/2,W/2]. Here $t=\hbar ^{2}/(2m^{\ast }a^{2})$ is the
hopping matrix element with the lattice constant $a$, $\mathbf{I}$ is a
two-dimensional identity matrix. The strength of Rashba SOI is represents by $%
V_{R}=\alpha \hbar /2a$, where $\alpha $ is the Rashba spin-orbital
coupling. $V_{R}$ is set to zero in the lead-2 and lead-4.

Considering a small temperature gradient $\Delta \mathcal{T}$ on the
longitudinal lead-1,3, we can set the temperatures $\mathcal{T}_{1}=\mathcal{%
T}+\Delta \mathcal{T}/2$, $\mathcal{T}_{3}=\mathcal{T}-\Delta \mathcal{T}/2$%
, $\mathcal{T}_{2}=\mathcal{T}_{4}=\mathcal{T}$. The charge current in lead-$%
p$ can be written as $J_{pe}=e(I_{p\uparrow }+I_{p\downarrow })$ and the
spin current is $J_{ps}=(\hbar /2)(I_{p\uparrow }-I_{p\downarrow })$. Here $%
I_{p\sigma }$ is the particle current in lead-$p$ with $\sigma $ equals to $%
\uparrow $ or $\downarrow $. $I_{p\sigma }$ can be obtained by the LB
formula:\cite{ref0,ref15}
\begin{equation}
I_{p\sigma }=\frac{1}{\hbar }\sum_{q\neq p}\int dE~T_{p\sigma
,q}(E)[f_{p}(E)-f_{q}(E)]  \label{Landau2}
\end{equation}%
where $T_{p\sigma ,q}(E)$ is the transmission coefficient from the lead-$q$
to the lead-$p$ with spin $\sigma $ and $E$ is the energy of the incident
electron. $f_{p}(E)$ is the electronic Fermi distribution function of the
lead-$p$.

The spin Hall current in lead-2 and lead-4 can be calculated with the closed
boundary condition in both lead-1,3 and lead-2,4, i.e. $V_{1}=V_{3}=0$ and $%
V_{2}=V_{4}=0$. From symmetry of the system, we know that $J_{2s}=-J_{4s}$%
\cite{ref0}. After the Taylor expansion, the spin Nernst coefficient $%
N_{s}\equiv J_{2s}/{\Delta \mathcal{T}}$ can be reduced to:%
\begin{equation}
N_{s}=\frac{1}{4\pi }\int {dE}(\Delta T_{23}-\Delta T_{21})\frac{E-E_{F}}{%
k_{B}\mathcal{T}^{2}}f(1-f),  \label{Ns}
\end{equation}%
here $\Delta T_{2p}=T_{2\uparrow ,p}-T_{2\downarrow ,p}$, and $f$ is the
zero order of Taylor expansion of the Fermi distribution function, it is the
same for all four leads, $f(E)=1/\{\mathrm{exp}[(E-E_{F})/k_{B}\mathcal{T}%
]+1\}$.
\begin{figure}[tbp]
\includegraphics[scale=0.72]{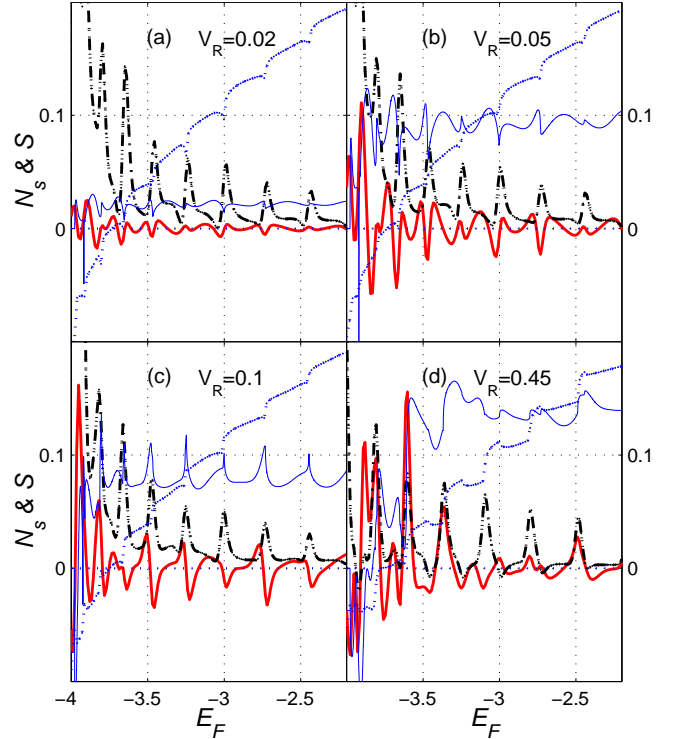}\newline
\caption{$N_{s}$ (red solid) and $S$ (black dotted) vs. Fermi energy $E_{F}$
for different Rashba $V_{R}$. The (scaled) transmission coefficient $%
T_{1,2}+T_{1,3}$ (thin blue dashed) and spin transmission coefficient $%
\Delta T_{2,3}$ (thin blue solid) are also shown. The other parameters are $%
\mathcal{T}=0.01$, and $L=19a$.}
\end{figure}

For the calculation of the longitudinal seebeck coefficient $S$, we need the
open boundary condition at lead-1,3, i.e. $J_{1e}=J_{3e}=0$ to find the
difference $\Delta V=V_{1}-V_{3}$. Different from a quasi-one-dimensional
2-leads system\cite{seebeck1}, the extra leads-2,4 also affects the
longitudinal seebeck coefficient $S$ of the entire system. For example, with
a perpendicular magnetic field $B$, the longitudinal seebeck coefficient $S$
is affected by the bias in leads-2,4, $V_{2}\ $and $V_{4}$. However, without
$B$, the sample's symmetry increases from $C_{2}$ symmetry to $D_{2}$
symmetry, i.e., we have $T_{1,2}=T_{1,4}$, Here $T_{1,2}=T_{1\uparrow ,2}+$ $%
T_{1\downarrow ,2}$. After the Taylor expansion, we can get the longitudinal
seebeck coefficient $S\equiv -\Delta V/\Delta \mathcal{T}$ as:%
\begin{equation}
S=\frac{1}{\mathcal{T}}\frac{\int dE~\left( T_{1,2}+T_{1,4}+2T_{1,3}\right)
(E-E_{F})f\left( 1-f\right) }{\int dE\left( T_{1,2}+T_{1,4}+2T_{1,3}\right)
f\left( 1-f\right) }.  \label{S}
\end{equation}%
The equation above shows that, even with a higher symmetry, the longitudinal
seebeck coefficient $S$ is still affected by the transport properties from
lead-2 and lead-4.

With the $D_{2}$ symmetry, the relationship between $S$ and $N_{s}$ can be
further derived. In fact, we can rewritten $S=\left( A^{\downarrow
}+A^{\uparrow }\right) \left/ \left[ \int dEF\left( \varepsilon \right)
\left( a^{\uparrow }+a^{\downarrow }\right) \right] \right. $. The $D_{2}$
symmetry gives $\Delta T_{23}=-\Delta T_{21}$. Noticing $T_{3\uparrow
,1}=T_{3\downarrow ,1}$, the spin Nernst coefficient can be simplified as $%
N_{s}\ =\left. \left( A^{\downarrow }-A^{\uparrow }\right) \right/ \left(
2\pi k_{B}\mathcal{T}\right) $. Here $\varepsilon =E-E_{F}$ and $F\left(
\varepsilon \right) =f\left( 1-f\right) $, $a^{\uparrow }$ denotes the spin
up term: $a^{\uparrow }=T_{2\uparrow ,1}+T_{3\uparrow ,1}$, and $%
a^{\downarrow }$ the spin down term $a^{\downarrow }=T_{2\downarrow
,1}+T_{3\downarrow ,1}$, we also use the notation of the integral term $%
A^{\uparrow }=\left. \int dE\varepsilon F\left( \varepsilon \right)
a^{\uparrow }\right/ \mathcal{T}$ and $A^{\downarrow }=\left. \int
dE\varepsilon F\left( \varepsilon \right) a^{\downarrow }\right/ \mathcal{T}$%
. Because of the symmetry, only leads-1,2,3 are used in the simplified
expression of $S$ and $N_{s}$, we only need the upper half of the sample for
our investigation. In fact, $\ a^{\uparrow }$ ($a^{\downarrow } $) and $%
A^{\uparrow }$ ($A^{\downarrow }$) reflects transport properties of spin-up
(spin-down) electrons in the upper half of the sample. Roughly speaking, $S$
can be seen as the sum of spin-up and spin-down terms, while $N_{s}\ $ as
the difference of them.

In the numerical calculations, $t=\hbar ^{2}/(2m^{\ast }a^{2})$ is set as
the energy unit. If taking the effective electron mass $m^{\ast }=0.05m_{e}$
and the lattice constant $a=12.5nm$, $t$ is about $5meV$. Temperature is
fixed by $k_{B}\mathcal{T}=0.01t$, which is about $1K$. The size of center
region is $L=19a$, about $237nm$. In a reasonable experimental range thus
far $V_{R}\in \left[ 0,0.1\right] $\cite{soi}. However, in order to
thoroughly study the relationship between the spin Nernst coefficient $N_{s}$
and the seebeck coefficient $S$, we extend the range of $V_{R}\ $up to $%
\left[ 0,1\right] $ in our calculation.
\begin{figure}[tbp]
\includegraphics[scale=0.6]{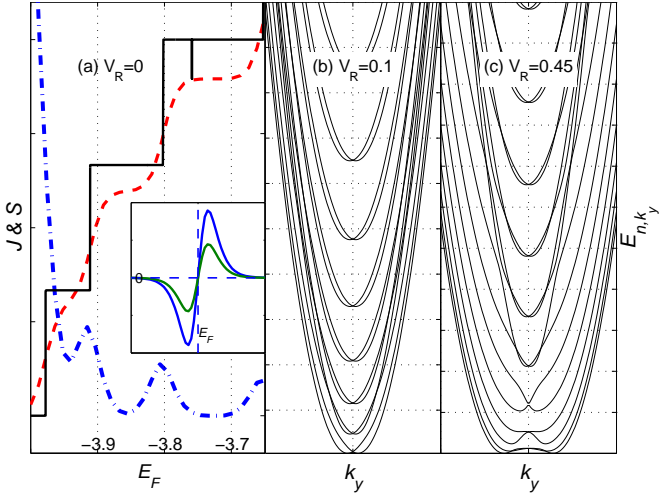}\newline
\caption{(a) A simple model: Current $J$ because of voltage gradient(red
dashed) and thermal power $S$ because of thermal gradient (blue dotted) vs.
Fermi energy $E_{F}$ at a two-lead system with Rashba $V_{R}=0$. The
(scaled) transmission function $T_{1,3}$ is also shown (black solid). The
plot in the small box shows $f_{L}-f_{R}$ with temperature difference. (b)
and (c): the eigen energy of the lead $E_{n,k_{y}}$ v.s. longitudinal wave
vector $k_{y}$ (units: $1/2a$) for different $V_{R}$}
\end{figure}

Fig.2 shows the spin Nernst coefficient $N_{s}$ and the seebeck coefficient $%
S$ versus the Fermi Energy $E_{F}$ in the clean system ($W=0$). It
is clearly seen that the seebeck coefficient $S$ peaks at the
positions where there are step-changes of transmission function
$T_{1,2}+T_{1,4}$. These peaks can be explained by a simple model
only with a 2-lead system without the Rashba SOI, shown in Fig.3(a).
The transmission coefficient $T_{1,3}$ is a step function
(solid-black curve). The reason is as follows. The sample can be
considered as a multi-channel system at a low temperature (here $T
\sim 1K$). When fermi energy increases, more channels in the lead
are used to transport current. Thus, $\Delta V$ of two leads as well
as the current increase with increasing of fermi energy (red-dashed
curve). However, the $S$ (blue-dotted curve) can not accumulates
while $E_F$ increases, it only peaks while the channel number
changes and $S$ is close to
zero with a fixed channel number. This can be seen from the LB formula (%
\ref{Landau2}), if lead-p and lead-q have different temperatures, $%
f_{p}(E,T+\Delta T)-f_{q}(E,T-\Delta T)$ is an antisymmetry function of $%
E-E_{F}$ (see plot in small box of Fig.3(a)): when $E<E_{F}$,
$f_{p}<f_{q}$, current flows from lower temperature lead to higher
temperature one; when $E>E_{F}$,
current flows in the opposite direction. Only when the two flows are not equal, i.e. $%
T_{p\sigma ,q}$ has an antisymmetry part, we can have a nonzero
current. Thus for Fig.3(a), only when $T_{1,3}$ is at the
step-change point, it has antisymmetry part and can give a non-zero
$S$.

\begin{figure}[tbp]
\includegraphics[scale=0.8]{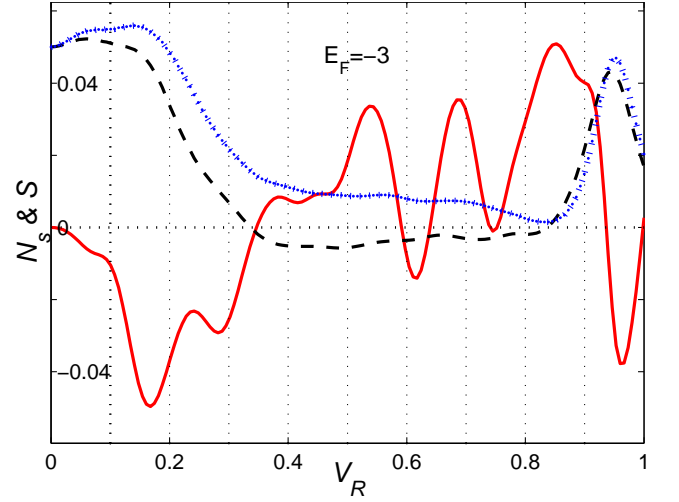}\newline
\caption{($N_{s}$ (red solid) and $S$ (black dashed) vs. Rashba SOI $V_{R}$
for fermi energy $E_{F}=-3$. For compare, the blue dotted line shows the
seebeck when lead-2,4 have the same $V_{R}$ as lead-1,3. The other
parameters are $\mathcal{T}=0.01$, and $L=19a$.}
\end{figure}

This conclusion can also be used to analyst spin-involved
quantities. From Fig.2, we can see that $N_{s}$ (red solid line)
shows an oscillatory structure. Besides the peaks at $V_{R}=0$
($N_{s}$ is zero at this point),
the magnitude of $N_{s}$ oscillation is also large at the peaks of $S$%
; but at the exact maximum point of $S$, where $V_{R}$ is quite small ($%
V_{R}\lesssim 0.1$), $N_{s}$ is generally close to zero. This is
because the spin transmission coefficient $\Delta T_{2,3}$ generally
has an extreme value when the transmission coefficient jumps at a
step. Around an extreme value, any function is almost symmetry, thus
one only can
get a low value of $N_{s}$. While at both sides of the extreme value, $%
\Delta T_{2,3}$ monotonically increases or decreases, we can get a
local maximum magnitude of $N_{s}$. Now why $\Delta T_{2,3}$ has an
extreme value at a peak of $S$ for a small $V_{R}$. Due to the
Rashba SOI, each eigen-energy band splits into two sub-bands with
opposite spin directions. These two sub-bands degenerate at
$k_{y}=0$, and the lower sub-band has two valleys below this
degenerate point. The two sub-bands are very close to each other
when $V_{R}$ is small. If the lower sub-band of high level (for
example, $E_{1,k_{y}}$) has the similar spin direction with the
upper
sub-band of low level energy (for example, $E_{0,k_{y}}$), $\Delta T_{2,3}$ continually increases when $%
E_{F}$ goes from the upper band of $E_{0,k_{y}}$ to the two valleys
of lower sub-band of $E_{1,k_{y}}$, and than rapidly decreases when
$E_{F}$ goes through the degenerate point ($k_{y}=0$) of
$E_{1,k_{y}}$, thus we get a peak in $\Delta T_{2,3}$; otherwise we
get a valley in $\Delta T_{2,3}$.

When $V_{R}$ is very big, we can see the external peaks for both
$N_{s}$ and $S$. For example $V_{R}=0.45$ in Fig.2(d), close to the
$2^{nd}$ and $3^{rd}$ main peaks of $S$, we can see a very sharp
sub-peak of $N_{s}$. In fact, these are also the small peaks of $S$,
though not very big. This is because for these two band (see
Fig.3(c)), the two valleys of the lower sub-band is far from the
degenerate point at $k_{y}=0$, the two channel of these two
sub-bands is separated. Thus we can see two peaks. At this time, the
change of spin transmission coefficients can be roughly thought as
the change of transmission coefficients, thus we can see $N_{s}$
peaks at the $S$'s peak.

\begin{figure}[tbp]
\includegraphics[scale=0.8]{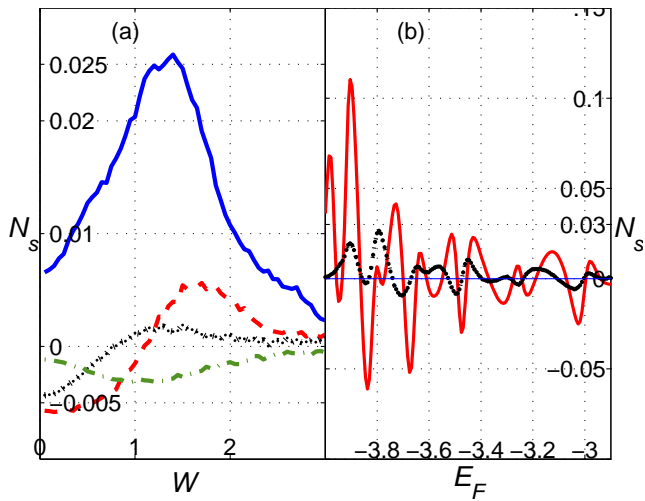}\newline
\caption{(a). $N_{s}$ vs. the strength of disorder $W$ for different
Fermi
level $E_{F}=-3.8$(solid blue), $E_{F}=-3.6$(dashed red), $E_{F}=-3.0$%
(dotted black), $E_{F}=-2.2$(dotted dashed green); (b). $N_{s}$ vs.
Fermi energy $E_{F}=-3.8$ for $W=0$ (dashed black) and $W=1.4$
(solid black).Other parameters are $V_{R}=0.05$, $\mathcal{T}=0.01$,
and $L=19a$.}
\end{figure}

In Fig.4, we show the spin Nernst coefficient $N_{s}$ and the
seebeck coefficient $S$ versus the Rashba SOI $V_{R}$ in the clean
system ($W=0$) for $E_{F}=-3$. The seebeck coefficient $S$ decreases
and maintains for a small value for quite a while before shows
another peak. This is because increasing $V_{R}$ moves the energy
bands and makes them go through the fermi energy. It should be
mentioned that we found the negative seebeck coefficient $S$ (Fig.
2d), which means a longitudinal current occurs in the opposite
direction of the temperature gradient $\Delta \mathcal{T}$. This is
due to the boundary conditions $V_{R}=0$ at leads-2,4. As a compare,
we also show $S$ for a uniform system, i.e. leads-2,4 having the
same strength of $V_{R}$ as in the sample. For this situation, the
seebeck coefficient $S$ is no longer negative. In fact, when the
Rashba SOI is absent in the leads-2,4, an interface between
$V_{R}=0$ and $V_{R}\neq 0$ ocurrs\cite{ref0}, this interface causes
additional scattering for an incident electron. In some special case
like $E_{F}=-3$, this may make the electrons below $E_{F}$ easier to
transport than the electrons above $E_{F}$, thus a negative $S$.

Finally we discuss the disorder effect on the spin Nernst effect. Fig.5
shows $N_{s}$ versus disorder strength $W$ for different Fermi energies. The
calculations are averaged over $500$ disorder configurations. Around $%
W<1.7$, $N_{s}$ shows an oscillatory structure. $N_{s}$ changes sign
with increasing of the disorder strength (see $E_{F}=-3.6$ and
$-3.0$ in Fig.5). It is interesting to see that, comparing to a
clean system ($W=0$), $N_{s}$ can be unexpectedly increased by
disorder $W$. This is because the disorder changes the oscillating
structure of $N_{s}$ (see Fig.5b). As expected, the disorder
decreases the strength of oscillating, however, it also shifts the
peak positions of $N_{s}$. It is possible to have a
peak in $N_s$ at finite disorder while it is almost zero initially at
clean limit. In Fig.5a, around $W=1\sim 1.5$, for the Fermi level $%
E_{F}=-3.8$, $-2.2$, we can see that $N_{s}$ is up to about three times of $%
N_{s}$ at $W=0$. The behavior of $N_{s}$ v.s. $W$ is very apparent when the
Fermi level $E_{F}$ is close to the bottom of energy band ($E_{F}=-4$). For $%
E_{F}=-3.8$, $N_{s}$ is much bigger than those at other Fermi
levels, and we can see a very remarkable peak at about $W=1.4$. For
$E_{F}=-2.2$, $N_{s}$ begins from $-0.005$, changes its sign at
about $W\sim 1.25$ and than increases, again reaches to $0.005$ at
about $W\sim 1.75$. With a very big disorder, $N_{s}$ should go to
zero as system enters into an
insulating regime. We find that the zero of $N_{s}$ occurs at $W=3$ for $%
V_{R}=0.05.$ This is roughly independent of the locations of the Fermi
energy.

In summary, in the absence of a perpendicular magnetic field, the
interplay between the spin Nernst effect and the seebeck effect is
investigated in a two-dimensional cross-bar with a spin-orbit
interaction. The spin Nernst effect exhibits an oscillatory
structure for a wide range of the Rashba SOI. With a large Rashba
SOI, the $N_s$ oscillation has a peak when the seebeck coefficient
possesses one. However, the inverse condition is not always
satisfied, namely, the seebeck coefficient can be almost zero while
$N_s$ has a peak. The disorder effect on the spin Nernst effects is
also studied. We find that disorder can enhance $N_{s}$ up to three
times for some Fermi levels. In addition, the disorder can also
change the sign of spin Nernst effect. Moreover, the limit of
disorder where $N_{s}$ goes to zero is independent of the Fermi
energy.

\textbf{Acknowledgments:} We thank Q.F. Sun and S.G. Cheng for many helpful
discussions. We gratefully acknowledge the financial support from US-DOE
under DE-FG02-04ER46124 and US-NSF.

\end{document}